\documentclass[aps,twocolumn,english,preprintnumbers,amsmath,amssymb,nofootinbib,superscriptaddress]{revtex4-1}
\usepackage{graphics,bm}
\usepackage{epsfig}
\usepackage{graphicx}
\usepackage{amsmath}
\usepackage{wrapfig}
\usepackage{bbm}
\usepackage{bm}
\usepackage{mathrsfs}
\usepackage{slashed}

\newcommand{\beq}{\begin{equation}}
\newcommand{\eeq}{\end{equation}}
\newcommand{\bqa}{\begin{eqnarray}}
\newcommand{\eqa}{\end{eqnarray}}

\def\square{\vcenter{\vbox{\hrule height.4pt
          \hbox{\vrule width.4pt height4pt
          \kern4pt\vrule width.3pt}\hrule height.4pt}}}

\providecommand{\f}[2]{\frac{{#1}}{{#2}}}

\newcommand{\ba}{\begin{eqnarray}}
\newcommand{\ea}{\end{eqnarray}}
\newcommand{\Hub}{\mathcal{H}}
\newcommand{\Mpl}{M_{\rm pl}}
\newcommand{\bphi}{\bar{\phi}}

\begin{document}


\title{Quantum corrections to inflaton dynamics, the semi-classical approach and the semi-classical limit}

\author{Matti Herranen}
\affiliation{Niels Bohr International Academy and Discovery Center, Niels Bohr Institute, Blegdamsvej 17, 2100 Copenhagen, Denmark}

\author{Asgeir Osland}
\author{Anders Tranberg}
\affiliation{Faculty of Science and Technology, University of Stavanger,
N-4036 Stavanger, Norway} 

\email{herranen@nbi.ku.dk}
\email{asgeir.osland@uis.no}
\email{anders.tranberg@uis.no}

\date{\today}

\begin{abstract}
Computations of quantum corrections to the CMB spectrum and to scalar field dynamics during inflation very often take advantage of the "semi-classical" approach, where the metric fluctuations are simply omitted. On the other hand, a complete computation ought to take into account that the matter field perturbation and scalar metric perturbation together constitute a single physical degree of freedom. The question then naturally arises, in which sense the semi-classical approach is an approximation to the complete calculation, and whether there are specific limits where this is also a good approximation. 
We demonstrate this by explicitly computing the leading quantum radiative corrections to the evolution equation of the mean field ("condensate") and the Friedmann equations taking into account scalar perturbations of both the matter field and the metric, and when omitting the latter. We find that the two agree in the limit $H\ll M_{\rm pl}$, but one is not a limit of the other. We also find that in simple models of inflation, $H/M_{\rm pl}$ is not small enough that the two approaches can be said to agree.
\end{abstract}

\keywords{cosmological perturbations, inflation, cosmology, quantum field theory in curved space-time}

\maketitle

\section{Introduction}
\label{sec:introduction}

The classical inflaton field\footnote{ Sometimes called the inflaton "condensate".}, often considered in generic models of inflation, should ultimately be identified with the expectation value (or mean field or one-point function) of some quantum scalar field. The dynamics of this "order parameter" (whether composite or fundamental) is then conveniently described in terms of the quantum effective action, from which the effective equations of motion arise through variation with respect to the field.

At tree level, this effective potential is the classical potential, and the standard text-book slow-roll treatment applies. This approach has been extremely successful in the interpretation of cosmological observations, in terms of different classical potentials \cite{Mukhanov,Watson,Bertschinger,Langlois,Malik,Baumann}. Beyond tree-level, quantum effects may however generate corrections to the Friedmann and scalar field equations. These can be significant, and in principle include effects from both the matter field fluctuations and metric perturbations. 
Because of gauge symmetry and constraints in the theory, metric and field scalar perturbations together represent only a single physical degree of freedom, and in different gauges, different variables may for convenience be chosen to vanish while others do not. 

We also know that at energy scales far below the Planck scale, where gravity effects are a priori negligible, one often simply ignores metric perturbations and proceeds to do perturbation theory in the matter field fluctuations only. This semi-classical approach to quantum corrections in curved space-time has a very long history, and provides an hugely popular formalism to compute corrections to the effective potential (such as recently in \cite{Higgspotential,tommi1,tommi2}), to the inflaton equation of motion (see for instance \cite{devega,Serreau:2011fu,Garbrecht1,Garbrecht2,Herranen:2013raa,Garbrecht:2011gu,Serreau:2013psa,Gautier:2013aoa}), and to the Friedman equations (for instance \cite{Herranen:2013raa}), by treating gravity as a classical background to a quantum field (for standard texts see \cite{birrell,Parker}). 

The standard computation of the CMB spectrum traditionally considers free fields, but when computing loop corrections to correlators, IR problems are encountered (divergences, secular behaviour, see for instance \cite{smitmeulen}). We know that these are unphysical, and will in the full theory be removed by the generation of effective masses. This is reminiscent of resummations in finite temperature field theory, where infrared divergences arise where the correct infrared physics has not been properly taken into account; in a sense the perturbative expansion has been carried out around an inappropriate vacuum (typically a massless propagator). In recent years, similar formalisms have been adapted to quantum fields in the context of inflation, and the IR problems were seen to indeed be unphysical and manageable (see for instance \cite{Weinberg,Sloth:2006az,Seery:2007we,Riotto:2008mv,Tranberg:2008ae,Senatore,Garbrecht:2011gu,Serreau:2011fu,Xue:2012wi,Serreau:2013psa,Gautier:2013aoa,Herranen:2013raa}). Most of these calculations were carried out in the semi-classical approach, where ambiguities around renormalisation can be readily resolved. 

If one were to include metric perturbations, it is possible that the perturbative non-renormalizability of gravity would jeopardise the resummed computation. If one could establish the semi-classical computation as a well-defined approximation to the metric-included computation (rather than an ad hoc approach), completely resummed and consistently renormalised semi-classical computations could be performed within the window of validity and directly used to describe the metric-included theory. 

Our aim in this paper is to compute the quantum corrected Friedmann and scalar field equations in both the semi-classical approach and when including the scalar metric perturbations from first principles, and show to what extent the former is an approximation to the latter. We emphasize that our immediate goal is not to reveal particular models that exhibit very large or observationally significant quantum corrections. We are concerned with establishing the relation between the complete result  and the semi-classical one. This may guide future considerations as to whether a given model is amenable to a semi-classical treatment. As an aside, to our knowledge computing the complete leading corrections to both the Friedmann and the scalar mean field equations has not been done before.

\subsection{"Semi-classical"}
\label{sec:semiclassical}

The label "semi-classical" has many uses in different areas of physics; we will adopt a specific terminology, designed to hopefully prevent confusion:
\begin{itemize}
\item The semi-classical {\it limit}, will be to take $H/M_{\rm pl}\rightarrow 0$, where $M_{\rm pl}$ is the Planck mass, and $H$ is the Hubble rate which determines the typical scale of matter field fluctuations in an inflationary background. This encodes the weakness of quantum gravity effects, and the rate of change of field mode frequencies and hence also adiabaticity. This is the ratio that is expected to suppress gravitational corrections relative to non-gravitational effects.
\item The semi-classical {\it approach} (SC) is in our terminology the choice of treating gravitational degrees of freedom classically (no fluctuations), and matter field degrees of freedom quantum mechanically. It is a priori distinct from the semi-classical limit. { The semi-classical approach also involves that for the Friedmann equations, one should take quantum expectation values of field correlators in the components of the energy momentum tensor, while leaving the background (FRW) metric unperturbed and classical.}
\item Closely related to the semi-classical {\it limit} is $H/k\rightarrow 0$, where $k$ is the momentum of a given field mode. These are the very sub-horizon modes at any given time, and are adiabatic since $H\ll\omega_k$. We will not give this limit a separate name.
\item The alternative to all of these is to neither use the semi-classical approach nor the semi-classical limit. This is then the {\it complete} quantum calculation involving fluctuations both in the matter fields and in the metric degrees of freedom. 
\end{itemize}
In broad terms, our task is then to investigate whether the semi-classical {\it approach} is in some way the semi-classical {\it limit} of the {\it complete} calculation, and quantify its range of validity.

The structure of the paper is as follows: In section \ref{sec:action}, we set our notation by deriving the tree level equations, specialising to Newtonian gauge. In section \ref{sec:fluctnewt} we quantise the constrained system of fluctuation equations, to leading order in slow-roll and compute a set of two-point vacuum correlators that will enter in the quantum corrected evolution equations. In section \ref{sec:qeom} we derive these equations. In section \ref{sec:semi-classical} we derive the analog equations in the semi-classical approach and discuss to what extent this approach can be seen as a limit of the complete calculation. For a particular set of models, we in section \ref{sec:examples} compute the actual magnitude of the quantum corrections. We conclude in section \ref{sec:conclusion}. A number of details are relegated to a series of short appendices (\ref{app:action}-\ref{app:SC_correlators}).

\section{Field dynamics in Newtonian gauge}
\label{sec:action}

We consider a single real, self interacting quantum scalar field $\phi$ evolving in a fluctuating background metric $g_{\mu\nu}$ close to a flat FRW Universe $\bar{g}_{\mu\nu}$. We can then write in a homogeneous state
\ba
\label{eq:fluct1}
\hat{\phi}(x)&=&\bar{\phi}(t)+\delta\phi({\bf x},t),\\
g_{\mu\nu}(x)&=&\bar{g}_{\mu\nu}(t)+\delta g_{\mu\nu}({\bf x},t),
\end{eqnarray}
with
\begin{eqnarray}
\delta g_{\mu\nu}&=&a^2h_{\mu\nu}=a^2\left[\begin{array}{cc}
2A&-B_{,i}\\
-B_{,i}&2(\psi\,\delta_{ij}-E_{,ij})\\
\end{array}\right],
\label{eq:pert_met}
\ea
and where $a(t)$ is the FRW scale factor, $\bar{\phi}=\langle\hat\phi\rangle$ is the mean field (or "classical" field), $\delta\phi$ is the field perturbation and $A$, $B$, $E$ and $\psi$ provide a parametrisation of the scalar degrees of freedom of the metric. Only one of these five degrees of freedom is physical. The rest can be removed by applying a gauge choice (two) and constraints arising from the action (two more). 

The action is
\ba
S=\int d^4x\sqrt{-g}\left[\frac{M_{\rm pl}^2}{2}R+\mathcal{L}_\phi(\phi,g_{\mu\nu})\right],
\label{orig_action}
\ea
where $M_{\rm pl}^2=(8\pi G)^{-1}$, $g$ is the determinant of the metric, $R$ is the Ricci scalar and $\mathcal{L}_\phi$ is the Lagrangian density of the scalar field, which we will specify in the following to be of the form
\ba
\mathcal{L}_\phi=\frac{1}{2}\partial_\mu\phi\,\partial^\mu\phi-V[\phi],
\ea
for some potential function $V[\phi]$.

We may then insert (\ref{eq:fluct1}) ff. into the action and keep terms to zeroth, first and second order in the fluctuations, so that
\ba
\label{eq:action_orders}
S=S^{(0)} + S^{(1)} + S^{(2)} + \mathcal{O}(\Delta^3) = S_{\rm g}+S_{\rm \phi}+\mathcal{O}(\Delta^3),\nonumber\\
\ea
where $\Delta$ denotes jointly $\{\delta\phi$, $A$, $B$, $E$, $\psi\}$. We find it convenient to use the Newtonian gauge
\ba
\label{gauge_condition}
E=B=0,
\ea
leaving only the $A$, $\psi$ and $\delta\phi$ variables.
 The explicit expressions for
 $S_{\phi,g}^{(0,1,2)}$ in the Newtonian gauge are found in Appendix \ref{app:action}. We adopt conformal time
 $\eta$, $dt=a(\eta)d\eta$, derivatives denoted by $'$.   

In order to derive the equations of motion for the gravitational field involving the mean metric $\bar{g}_{\mu\nu}$ (parametrized by the scale factor $a(\eta)$ for a flat FRW Universe) and the
 fluctuation $\delta g_{\mu\nu}$ (parametrized by $A$, $\psi$) we first take a variation of the gauge
 un-fixed action (\ref{orig_action}) with respect to the full metric $g_{\mu\nu}$ to obtain the Einstein
 equations. We then expand these to second order in fluctuations $\Delta$ in Newtonian gauge,
 and finally take the quantum expectation value to extract the equations of motion for the fluctuations
 and the mean fields, given below by Eqs.~(\ref{eq:fried1}-\ref{eq:fried4}) and (\ref{eq:Qfried1}-\ref{eq:Qfried2}), respectively. A direct variation of the
 gauge-fixed action (\ref{action_app}-\ref{last_eq}) with respect to $a$, $A$ and $\psi$ would give
 an incorrect result for the equations of motion.

On the other hand, the corresponding equations of
 motion for the scalar field fluctuation $\delta\phi$ and the mean field $\bar\phi$, given below by
 Eqs.~(\ref{eq:fluct}) and (\ref{eq:phieom}), respectively, can be derived either by a direct variation of
 the gauge-fixed action (\ref{action_app}-\ref{last_eq}) with respect to $\delta\phi$ and $\bar\phi$,
 or by a variation of the gauge un-fixed action (\ref{orig_action}) and subsequently fixing the gauge
 and taking the quantum expectation value. The reason why both procedures work, i.e. gauge fixing and
 variation of action commute, for the scalar field but not for the gravitational field is that the gauge
 fixing condition (\ref{gauge_condition}) involves the gravitational degrees of freedom $E$ and $B$ but
 not the scalar field $\phi$. 
 Further details on the derivation of the equations of motion are presented in Appendix \ref{app:eom}.

\subsection{Classical equations of motion: Tree level}
\label{sec:treelevel}

At tree level, all the fluctuations $\delta\phi$, $A$, $B$, $E$, $\psi$ are set to zero. Variation of $S_0$ gives us the classical or tree level field equation of motion and Friedmann equations. 
\ba
\label{eq:tree}
0&=&\bphi''+2\Hub\bphi'+V_{,\phi}[\bphi],\\
\label{eq:tree2}
3\Mpl^2\Hub^2&=&\frac{1}{2}\bphi'^2+a^2V[\bphi],\\
3\Mpl^2\Hub'&=&-\bphi'^2+a^2V[\bphi],
\label{eq:tree3}
\ea
where we have defined the "conformal" Hubble rate $\Hub=a'/a = aH$. The task is to find the quantum corrected versions of (\ref{eq:tree}-\ref{eq:tree3}).

The tree level equations form the basis of the slow-roll expansion (SR), where defining
\ba
\label{SR-parameters}
\epsilon = -\frac{\dot{H}}{H^2}=1-\frac{\Hub'}{\Hub^2},
\qquad \delta = -\frac{\ddot{\bphi}}{H\dot{\bphi}}=1-\frac{\bphi''}{\Hub\bphi'}\nonumber\\
\ea
we have
\ba
\label{eq:hubeps}
\Hub=-\frac{1}{\eta}(1+\epsilon)+\mathcal{O}(\epsilon^2),\qquad
\epsilon\Hub^2 = \frac{\bphi'^2}{2\Mpl^2}=4\pi G\,\bphi'^2.\nonumber\\
\ea
For the more general quantum corrected mean field and Friedmann equations, standard slow-roll manipulations are not exact, and these definitions are only approximately applicable. We will however adopt them in the following, and rank terms in powers  of $\epsilon$ and $\delta$.

\section{Fluctuations}
\label{sec:fluctnewt}

We now return to the action $S^{(0)} + S^{(1)} + S^{(2)}$ (\ref{eq:action_orders}). By variation w.r.t. the metric, and inserting the tree level relations (\ref{eq:tree}-\ref{eq:tree3}), we find for the first order Einstein equations 
\ba
\label{eq:fried1}
\nabla^2\psi-3\Hub(\psi'+\Hub A)&\\=4\pi G &(-A\bphi'^2+a^2V_{,\phi}\delta\phi+\bphi'\delta\phi'),\nonumber\\
\label{eq:fried2}
\psi''+(\Hub^2+2\Hub')A+\Hub (A'+2\psi')&\\=-4\pi G&(A\bphi'^2+a^2V_{,\phi}\delta\phi-\bphi'\delta\phi'),\nonumber\\
\label{eq:fried3}
\psi'+\Hub A=4\pi G\bphi'\delta\phi,\\
\label{eq:fried4}
A=\psi.
\ea
The last equation follows in Newtonian gauge from assuming the absence of anisotropic stress (see Appendix \ref{app:eom}). When including the equation of motion for the scalar field fluctuations, 
\ba
\label{eq:fluct}
\delta\phi''+2\Hub\delta\phi'-\nabla^2\delta\phi+a^2V_{,\bar{\phi}\bar{\phi}}\delta\phi&\nonumber\\=4\bar{\phi}'A'&+(4\bar{\phi}''+4\Hub\bar{\phi}'-2V_{,\bar{\phi}}) A,\nonumber\\
\ea
one of the equations becomes redundant. 
Quantizing a constrained system is a standard procedure, in terms of the canonical momenta 
\ba
\label{eq:cons1}
\Pi_A&\equiv& \frac{\partial\mathcal{L}}{\partial A'}=0,\\
\label{eq:cons2}
\Pi_\psi&\equiv& \frac{\partial\mathcal{L}}{\partial \psi'}=-\frac{3a^2}{4\pi G}\left[\psi'+\Hub(A+\psi)\right],\\
\label{eq:cons3}
\Pi_{(\delta\phi)}&\equiv& \frac{\partial\mathcal{L}}{\partial (\delta\phi')}=a^2\left[\delta\phi'-\bar{\phi}'(A+3\psi-1)\right],
\ea
Eq.~(\ref{eq:cons1}) is a primary constraint for the (auxiliary) field $A$, while the corresponding
 equation of motion (\ref{eq:fried1}) provides a secondary constraint. Eqs.~(\ref{eq:fried3}, \ref{eq:fried4})
 provide two additional constraints, so in total we have four constraints for the canonical variables
 $\psi$, $A$, $\delta\phi$ and their conjugate momenta:
\begin{align}
\chi_i = 0\,,\qquad i=1,\ldots,4\,,
\end{align}
with
\begin{align}
\chi_1 \equiv &\Pi_A,
\label{constr1}
\\
\chi_2 \equiv &\nabla^2\psi + 3\mathcal{H}'\psi -\nonumber\\
 4\pi G& \left( a^2 V_{,\phi}\delta\phi
 + a^{-2}\bphi'\,\Pi_{\delta\phi} - a^{-2}\mathcal{H}\,\Pi_\psi
 - \bphi'^2\right),
\label{constr2}
\\
\chi_3 \equiv &\mathcal{H}\psi + 4\pi G\Big(\bphi'\delta\phi + \frac{1}{3a^2}\Pi_\psi \Big),
\label{constr3}
\\
\chi_4 \equiv &A - \psi\,,
\label{constr4}
\end{align}
where we have used the zeroth order background equations and Eqs.~(\ref{eq:cons2}, \ref{eq:cons3}) to solve
 the time-derivatives for the canonical momenta. This leaves 6 - 4 = 2 variables or one physical degree of freedom, as expected.

\subsection{Constrained quantisation}

The auxiliary field $A$ and its conjugate momentum $\Pi_A$ can be readily solved for by using the constraints
 $\chi_1$ and $\chi_4$. The remaining constraints $\chi_2$ and $\chi_3$ are of the {\em second class} since
 their Poisson bracket $[\chi_2, \chi_3]_P$ does not vanish. To quantize the variables $\psi$ and
 $\delta\phi$ subject to second class constraints $\chi_{2,3}$ we define the constraint matrix:
\begin{align}
C_{mn} = [\chi_m, \chi_n]_P\,,\qquad m,n = 2,3 \,,
\label{constr_matrix}
\end{align}
and the Dirac brackets:  
\begin{align}
[A,B]_D \equiv [A,B]_P \;- \sum_{m,n=2,3}[A,\chi_m]_P \big(C^{-1}\big)_{mn}[\chi_n,B]_P\,,
\end{align}
where the standard Poisson bracket is defined as ($\psi_a = \{\psi, \delta\phi\}$)
\begin{align}
[A,B]_P \equiv \sum_a \left(\frac{\partial A}{\partial\psi_a}\frac{\partial B}{\partial\Pi_{\psi_a}}
-\frac{\partial B}{\partial\psi_a}\frac{\partial A}{\partial\Pi_{\psi_a}}\right)\,.
\end{align}
The equal-time commutation relations of quantized variables are then given by:
\begin{align}
[A,B] \equiv i[A,B]_D.
\label{equal-time_comm_def}
\end{align}
The resulting commutation relations for $\psi$, $\delta\phi$ and their conjugate momenta $\Pi_\psi$,
 $\Pi_{\delta\phi}$ are presented in Appendix \ref{app:comm_relations}. 
By using the relations (\ref{eq:cons2}, \ref{eq:cons3}) we can solve for the conjugate momenta in terms of
 time derivatives to find
\begin{align}
&[\psi(\mathbf{x}),\psi(\mathbf{y})] = [\delta\phi(\mathbf{x}),\delta\phi(\mathbf{y})] = 0,
\label{commutation_psi2a}
\\[3mm]
&[\psi(\mathbf{x}),\psi'(\mathbf{y})] = -i\frac{\big(4 \pi G \bphi'\big)^2}
{a^2 \nabla_x^2}\delta^3(\mathbf{x}-\mathbf{y})\,,
\label{commutation_psi2}
\\[2mm]
&[\delta\phi(\mathbf{x}),\delta\phi'(\mathbf{y})]
 = \frac{i}{a^2}\left(1 + \frac{4 \pi G \bphi'^2}{\nabla_x^2}\right)
\delta^3(\mathbf{x}-\mathbf{y})\,,
\label{commutation_dphi2}
\end{align}
where we have suppressed the (equal) time arguments of the fields. Remembering Eq. (\ref{eq:hubeps}), we see that the usual flat-space commutation relation is recovered from Eq.~(\ref{commutation_dphi2}) in the $\epsilon\rightarrow 0$ limit, as well as for modes well within the horizon $\Hub/{\bf |k|}\rightarrow 0$. { The $\epsilon\rightarrow 0$ limit is singular in the sense that (\ref{commutation_psi2}) vanishes.}

This procedure and the commutation relations (\ref{commutation_psi2a}, \ref{commutation_psi2}) are consistent with treating $\psi$ as the only dynamical d.o.f. while $\delta\phi$ and $A$ are expressed in terms of $\psi$ using the constraints (\ref{eq:fried3}, \ref{eq:fried4}). In this way the dynamical equations (\ref{eq:fried1}, \ref{eq:fried2}) can be written as:
\ba
\psi''-\nabla^2\psi+2\left(\Hub-\frac{\bphi''}{\phi'}\right)\psi'+2\left(\Hub'-\frac{\bphi''}{\bphi'}\Hub\right)\psi=0.\nonumber\\
\ea
{ We note that by substitution or in other gauges, the variable $\delta\phi$ or $A$ may appear with a second time derivative, and hence be the dynamical variable (and indeed, $\psi=A$ here). In the present case, however, the mode equations are simplest in terms of $\psi$.}

\subsection{Mode functions}
\label{sec:mode}

The field $\psi$ can be decomposed in terms of the mode functions $f_\mathbf{k}(\eta)$ as
\begin{align}
\label{field_operator}
\psi(\eta,\mathbf{x}) =  \int \frac{d^3 k}{(2\pi)^3}
\left[a_\mathbf{k} f_\mathbf{k}(\eta) e^{i \mathbf{k \cdot x}} + a_\mathbf{k}^{\dagger} f_\mathbf{k}^*(\eta)
 e^{-i \mathbf{k \cdot x}}\right]\,,
\end{align}
where the creation and annihilation operators satisfy the standard commutation relations
\begin{align}
[\hat{a}_\mathbf{k},\hat{a}_{\mathbf{k}'}]=[{\hat{a}}^\dagger_\mathbf{k},{\hat{a}}^\dagger_{\mathbf{k}'}]=0,
\qquad[{\hat{a}}_\mathbf{k},{\hat{a}}^\dagger_{\mathbf{k}'}]=(2\pi)^3 \delta^3(\mathbf{k}-\mathbf{k}')\,,
\label{eq_com}
\end{align}
and the mode functions satisfy the equation
\begin{align}
\label{mode_eq}
f_\mathbf{k}'' + 2\left(\mathcal{H}
-\frac{\bphi''}{\bphi'}\right)
f_\mathbf{k}'+2\left(\mathcal{H}'-\frac{\bphi''}{\bphi'}\mathcal{H}\right)f_\mathbf{k}
 + |\mathbf{k}|^2 f_\mathbf{k} = 0,
\end{align}
as well as the Wronskian which fixes the normalization\footnote{Note that the scaled mode functions
 $\tilde f_\mathbf{k} = a|\mathbf{k}|/(4\pi G \bphi')f_\mathbf{k}$ satisfy the standard Wronskian
$\tilde f_\mathbf{k} \tilde f_\mathbf{k}'^* - \tilde f_\mathbf{k}^* \tilde f_\mathbf{k}' = i$.}
\begin{align}
\label{wronskian}
f_\mathbf{k}(\eta)f_\mathbf{k}'^*(\eta) - f_\mathbf{k}^*(\eta)f_\mathbf{k}'(\eta)
 = i\left(\frac{4\pi G \bphi'}{a|\mathbf{k}|}\right)^2,
\end{align}
in order to accommodate the equal-time commutation relations (\ref{commutation_psi2}). Using
 Eq.~(\ref{SR-parameters}) the mode equation (\ref{mode_eq}) can be written to first order
 in the slow-roll expansion as
\begin{align}
f_\mathbf{k}'' - \frac{2\delta}{\eta}f_\mathbf{k}' + \left[\frac{2(\delta - \epsilon)}{\eta^2}
 + |\mathbf{k}|^2 \right]f_\mathbf{k} =0\,,
\end{align}
with the solution satisfying the constraint (\ref{wronskian}) given by\footnote{We neglect the
 time-dependence of the slow-roll parameters $\epsilon$ and $\delta$, which is parametrically
 second order in slow roll.}
\begin{align}
\label{eq:vacmode}
f_\mathbf{k}(\eta) = \sqrt{\frac{\pi\epsilon}{8}}\frac{\Hub}{M_p a|\mathbf{k}|}
(-\eta)^{1/2}H^{(2)}_\nu(k\eta)\,,
\nonumber\\
\nu = \frac{1}{2}\sqrt{1+8\epsilon - 4\delta} \approx \frac{1}{2} + 2\epsilon - \delta\,.
\end{align}
%

\subsection{Vacuum correlators}
\label{sec:loop}

In the scale factor and mean field equations, the quantum corrections will appear in terms of two-point
 correlators, with up to four time- or spatial derivatives. Having solved for the quantum slow-roll
 vacuum mode functions (\ref{eq:vacmode}), these can now be computed explicitly. 
 Using Eqs.~(\ref{field_operator}, \ref{eq_com}) we get for the loop contribution in the vacuum state:
\begin{align}
\label{loop_text}
\langle \psi^2 \rangle  =  \int \frac{d^3 k}{(2\pi)^3} |f_\mathbf{k}|^2
 = \frac{\epsilon \Hub^2}{16\pi M_p^2 a^2}
\int_{\Lambda_{\rm IR}}^{\Lambda_{\rm UV}} dx \big|H^{(2)}_{\nu}(-x)\big|^2\,,
\end{align}
where we use (constant) UV cutoff $\bar\Lambda_{\rm UV}$ for the physical momenta: 
$k/a = -kH\eta < \bar\Lambda_{\rm UV}$ such that $x = -k \eta < \bar\Lambda_{\rm UV} / H \equiv \Lambda_{\rm UV}$
 and a similar IR cutoff $\Lambda_{\rm IR}$, which is related to the duration of inflation below.
 The details of evaluating the loop integrals for
 $\langle \psi^2 \rangle$, $\big\langle \psi\nabla^2 \psi \big\rangle$ and
 $\big\langle \psi\nabla^4 \psi \big\rangle$ are presented in Appendix \ref{app:loop_integrals}.
 The results are given by
\begin{align}
\label{loop_final}
C_0 \equiv \langle \psi^2 \rangle = &\frac{\epsilon \Hub^2}{8\pi^2 M_p^2 a^2}
\bigg[\bigg(\frac{1}{2(\delta-2\epsilon)} - \frac{1}{2} + \log 2 + \gamma_E\bigg)\times\nonumber\\&
\bigg(1 - \Lambda_{\rm IR}^{2\delta - 4\epsilon}\bigg) + \log\Lambda_{\rm UV}\bigg]
+ \ldots
\\[2mm]
\label{k2loop_final}
C_2 \equiv-\big\langle \psi\nabla^2 \psi \big\rangle
=& \frac{\epsilon \Hub^2}{16\pi^2 M_p^2 a^2}(-\eta)^{-2}
\Big[\Lambda_{\rm UV}^2 \nonumber\\&-(\delta - 2\epsilon)\log\Lambda_{\rm UV}
 - \Lambda_{\rm IR}^2
\Big] + \ldots
\\[3mm]
\label{k4loop_final}
C_4 \equiv \big\langle \psi\nabla^4 \psi \big\rangle
= &\frac{\epsilon \Hub^2}{32\pi^2 M_p^2 a^2}(-\eta)^{-4}
\Big[\Lambda_{\rm UV}^4 - (\delta - 2\epsilon)\Lambda_{\rm UV}^2\nonumber\\&
+ \frac{1}{16}(\delta - 2\epsilon)^2 (2 + \delta - 2\epsilon)^2\log\Lambda_{\rm UV}
- \Lambda_{\rm IR}^4
\Big] \nonumber\\&+ \ldots\,,
\end{align}
where we have neglected the (finite) terms of order ${\cal O}(\epsilon,\delta)$ and higher\footnote{For
 the UV divergences we formally keep also higher order terms in $\epsilon$ and $\delta$, which can be computed
 reliably.} while the positive powers of the IR cutoff $\Lambda_{\rm IR} \ll 1$ will be neglected below. 
The other two-point correlators of relevance involving derivatives of the fields can be related to the
 expressions (\ref{loop_final}-\ref{k4loop_final}) by using the mode equation (\ref{mode_eq}). The
 resulting relations are given by Eqs.~(\ref{correlator_relations}). 

\subsection{IR cutoff and the duration of inflation}

We will now briefly discuss how the IR cutoff $\Lambda_{\rm IR}$ can be related to the duration of 
 inflation. A natural prescription, following \cite{Xue:2012wi}, is to set the IR cutoff such that the
 modes that were superhorizon already at the beginning of the inflation (and hence throughout inflation)
 do not contribute to the loop integrals. This means that the comoving momenta are cut off by the initial
 Hubble radius:
\begin{equation}
k \geq a_{\rm in}H_{\rm in}\,,
\end{equation}
which leads to $x \geq \Lambda_{\rm IR}$ with
\begin{equation}
\Lambda_{\rm IR} = \f{a_{\rm in}H_{\rm in}}{aH(1-\epsilon)}\,.
\label{cutoff_rel1}
\end{equation}
For approximately constant $\epsilon$, we find
\begin{equation}
a = a_{\rm in}e^N\,,\qquad\qquad
H = H_{\rm in}e^{-\epsilon N}\,,
\label{N-dep}
\end{equation}
where 
\begin{equation}
N(t) \equiv \int_{t_{\rm in}}^t dt' H(t'),
\end{equation}
is the number of $e$-foldings from the beginning of the inflation. Using these expressions we now get for
 the cutoff in (\ref{cutoff_rel1}) 
\begin{equation}
\Lambda_{\rm IR} = \f{1}{1-\epsilon}e^{-(1-\epsilon)N}\,,
\label{cutoff_rel2}
\end{equation}
such that the logarithm of $\Lambda_{\rm IR} \ll 1$ is approximately given by
\begin{equation}
|\log\Lambda_{\rm IR}| \approx (1-\epsilon)N\,.
\label{cutoff-N_rel}
\end{equation}
%

\subsection{Renormalization and IR behaviour}
\label{sec:renorm}

The correlators $C_{0,2,4}$ are UV-divergent, and in order to have sensible equations of motion, we need to introduce a renormalisation prescription. We have used a simple cut-off regularisation which explicitly breaks Lorentz symmetry, and ideally one would wish to redo the computation in a general number of spatial dimensions, making use of dimensional regularisation. In that case, renormalisation may be recast in terms of counterterms for invariant operators $R$, $R^2$, $R^{\mu\nu}R_{\mu\nu}$ (see for instance \cite{tommi1,tommi2}). For our purposes here, it will be sufficient to adopt a MS-like prescription, whereby all UV-divergent terms are subtracted and hence can be discarded in the following. 

Doing this, and keeping in addition only terms to leading order in slow-roll, we find
\begin{align}
C_0 &=\frac{\epsilon \Hub^2}{16\pi^2M^2_{\rm pl} a^2 (\delta-2\epsilon)}
\bigg(1 - \Lambda_{\rm IR}^{2\delta - 4\epsilon}\bigg)+ \ldots \,,\\
C_2&=-\frac{\epsilon \Hub^4}{16\pi^2M^2_{\rm pl} a^2}\Lambda_{\rm IR}^2\simeq 0\,,
\\
C_4 &=-\frac{\epsilon \Hub^6}{32\pi^2M^2_{\rm pl}a^2}\Lambda_{\rm IR}^4\simeq 0\,,
\end{align}
where for $C_0$ we have written down only the dominant contribution in the slow roll expansion. $C_0$
 involves a logarithmic IR divergence in the limit $\Lambda_{\rm IR} \to 0$, while the positive
 powers of $\Lambda_{\rm IR}$ in $C_2$ and $C_4$ give negligible contributions for very small cutoff
 ($\Lambda_{\rm IR}\ll 1$). When the logarithm of $\Lambda_{\rm IR}$ is related to number of $e$-foldings
 via (\ref{cutoff-N_rel}), we find that $C_0$ can be expanded in two limits:
 $N \ll N_{\rm sat}$ and $N \gg N_{\rm sat}$, with
\begin{equation}
\label{Nsat}
N_{\rm sat} \equiv \frac{1}{|2\delta - 4 \epsilon|}\,,
\end{equation}
to find
\begin{equation}
\label{eq:C024}
C_0 = \frac{\epsilon \Hub^2}{8\pi^2M^2_{\rm pl} a^2}N_{\rm eff}\,,
\end{equation}
where
\begin{eqnarray}
\label{Neff}
N_{\rm eff} &\equiv& \frac{1}{2\delta-4\epsilon}
\bigg(1 - e^{-(2\delta - 4\epsilon)|\log\Lambda_{\rm IR}|}\bigg)\nonumber\\ 
&\simeq& \left\{ \begin{array}{ll}
N\,, & \qquad N \ll N_{\rm sat}\,,\\
N_{\rm sat}\,, & \qquad N \gg N_{\rm sat}\,,\quad\delta > 2\epsilon\,,\\
N_{\rm sat}e^{N/N_{\rm sat}}\,, & \qquad N \gg N_{\rm sat}\,,\quad\delta < 2\epsilon\,.
\end{array}\right.\nonumber\\
\end{eqnarray}
We see that the loop
 contribution $C_0$ grows linearly for small $N$ and after $N \sim N_{\rm sat}$ $e$-foldings it
 either saturates to a value proportional to $N_{\rm sat}$ for $\delta > 2 \epsilon$ or grows exponentially
 for large $N$ in the case of $\delta < 2 \epsilon$ signalling the breakdown of perturbative loop expansion
 and calling for non-perturbative resummation. The behavior of the loop contributions with linear growth
 and subsequent saturation as a function of $N$ is in qualitative agreement with similar studies using the
 non-perturbative stochastic approximation \cite{Riotto:2008mv,Starobinsky1,Starobinsky2,Garbrecht1,Garbrecht2}, although for the latter the expression of $N_{\rm sat}$ depends on the model and interactions. 

Moreover, we can relate the denominator in (\ref{Nsat}-\ref{Neff}) to the scalar spectral index at the horizon
 crossing, defined below in (\ref{spectrum}), to get:
\ba
\label{spectrum_rel}
2\delta - 4\epsilon = \frac{2}{3}\delta_M - 6\epsilon = n_s - 1\,,
\ea 
and therefore we see that for the observed spectral index at $k_* = 0.05\;{\rm Mpc}^{-1}$\cite{Ade:2015oja}: $n_s = 0.9655 \pm 0.0062$
 (68\% CL, Planck TT + lowP), $\delta < 2\epsilon$ and hence for realistic inflationary parameters the perturbative
 expansion indeed appears to break down for large $N \gtrsim N_{\rm sat}$. Below in Section \ref{sec:examples} we estimate
 the size of quantum radiative corrections for realistic inflation models by assuming that the true non-perturbative
 saturation value for $N_{\rm eff}$ would be of order $N_{\rm sat}$.

As we will see below, also derivatives of the loop contributions with respect to conformal time enter the quantum corrected equations of motion,
 but we have to leading order in slow-roll
\ba
&\partial_\eta C_0 = \partial_\eta^2C_0= \partial_\eta C_2 = \partial_\eta^2 C_2 = 0+\mathcal{O}(\epsilon^2).
\ea

\section{Quantum corrected equations of motion}
\label{sec:qeom}

The derivation of the quantum corrected equations of motion is discussed in detail in Appendix B. For the mean field $\bar\phi$ we find the evolution equation
\ba
&&\bphi''\bigg[1+\frac{1}{\epsilon}\bigg(
(8+6\epsilon)C_0+6\frac{\partial_\eta C_0}{\Hub}+2\frac{\partial^2_\eta C_0}{\Hub^2}+4\frac{C_2}{\Hub^2}
\bigg)\bigg]
\nonumber\\&&
+2\Hub\bphi'\bigg[1+
\frac{1}{\epsilon}\bigg((-4-\delta_M+10\delta
)C_0
\nonumber\\&&
\qquad\qquad-\frac{1}{2}(\delta_M+6-2\epsilon-14\delta
)\frac{\partial_\eta C_0}{\Hub}
-(1-2\delta)\frac{\partial^2_\eta C_0}{\Hub^2}
\nonumber\\&&
\qquad\qquad-4\frac{C_2}{\Hub^2}-2\frac{\partial_\eta C_2}{\Hub^3}
\bigg)\bigg]+a^2V_{,\phi}\left[1-2C_0\right]
=\nonumber\\&&
-\frac{M_{\rm pl}^2}{\epsilon}a^2V_{,\phi\phi\phi}[\bphi]
\bigg[
(1-2\epsilon+2\delta)C_0
\nonumber\\&&
\qquad\qquad\qquad\qquad+(1+\delta)\frac{\partial_\eta C_0}{\Hub}+\frac{1}{2}\frac{\partial^2_\eta C_0}{\Hub^2}
\bigg].\label{eq:phieom}
\ea
We have introduced the quantity
\ba
\label{eq:deltaM}
\delta_M\equiv\frac{a^2V_{,\phi\phi}[\bphi]}{\Hub^2}\simeq 3(\delta+\epsilon),
\ea
where the last relation is correct at leading order in SR. We may therefore take $\delta_M$ to be small whenever the SR parameters are.
Similarly, we have the quantum corrected Friedmann equations
\ba
\label{eq:Qfried1}
&&3M_{\rm pl}^2\Hub^2=\left[\frac{1}{2}\bphi'^2+a^2V[\bphi]\right]
+\frac{M_{\rm pl}^2\Hub^2}{\epsilon}\bigg[
-12\epsilon
C_0
-\frac{3}{2}
\epsilon
\frac{\partial_\eta^2 C_0}{\Hub^2}\nonumber\\&&
-(1+5\epsilon+4\delta
)\frac{C_2}{\Hub^2}
-(1+2\delta)\frac{\partial_\eta C_2}{\Hub^3}
-\frac{1}{2}\frac{\partial_\eta^2 C_2}{\Hub^4}+2\frac{C_4}{\Hub^4}
\nonumber\\&&
+
\delta_M\bigg(
(1-2\epsilon+2\delta)C_0+(1+\delta)\frac{\partial_\eta C_0}{\Hub}+
\frac{1}{2}\frac{\partial_\eta^2C_0}{\Hub^2}-\frac{C_2}{\Hub^2}
\bigg)
\bigg]
,\nonumber\\
\\
\label{eq:Qfried2}
&&3M_{\rm pl}^2\Hub'=\left[-\bphi'^2+a^2V[\bphi]\right]
-\frac{M_{\rm pl}^2\Hub^2}{\epsilon}\bigg[
(12\epsilon-8\delta
)C_0
\nonumber\\&&
+(10\epsilon-8\delta
)\frac{\partial_\eta C_0}{\Hub}
+(3-8\delta
)\frac{\partial_\eta^2 C_0}{\Hub^2}
+(4-12\epsilon+4\delta
)\frac{C_2}{\Hub^2}
\nonumber\\&&
+(4-6\delta)\frac{\partial_\eta C_2}{\Hub^3}
+2\frac{C_4}{\Hub^4}
\nonumber\\&&
-\delta_M\bigg(
(1-2\epsilon+2\delta)C_0+(1+\delta)\frac{\partial_\eta C_0}{\Hub}+
\frac{1}{2}\frac{\partial_\eta^2C_0}{\Hub^2}-\frac{C_2}{\Hub^2}\bigg)\bigg].\nonumber\\
\ea
As advertised, all the quantum corrections are in terms of the correlators considered in section \ref{sec:fluctnewt}, and computed in Appendices \ref{sec:correlators} and \ref{app:loop_integrals}. It is clear that removing all the correlators, we recover the classical result from section \ref{sec:treelevel}. We have introduced the slow-roll parameters wherever possible, keeping the lowest and next-to-lowest order in slow-roll. Given a general solution to the mode functions, to whatever order of interest, the correlators $C_{0,2,4}$ can now be inserted. The mode functions we found in section \ref{sec:mode} are only accurate to first order in slow-roll, and inserting these consistently, we would have to 
 discard all terms of higher order also in the resulting equations of motion.
We will do so explicitly below, but since all the correlators have an overall factor of $\epsilon \Hub^2/M_{\rm p}^2$, the leading term will be of the type $C_0/\epsilon$. 

What is also important to note, is that we should not expect this computation in Newtonian gauge to agree with computations in other gauges. The mode functions and therefore $C_{0,2,4}$ will also be different in different calculations, and only when inserting these should the results agree. 

\subsection{Leading order in slow-roll}

To leading order in slow-roll the correlators ($\mathcal{O}(\epsilon,\delta)$)  combine with the equations of motion to give quantum corrections at $\mathcal{O}(1)$. We have for the Friedmann equations 
\ba
3M_{\rm pl}^2\Hub^2&=&\left[\frac{1}{2}\bphi'^2+a^2V[\bphi]\right]
+\frac{M_{\rm pl}^2\Hub^2}{\epsilon}(\delta_M-12\epsilon)
C_0^{(\epsilon)},
\\
3M_{\rm pl}^2\Hub'&=&\left[-\bphi'^2+a^2V[\bphi]\right]
+\frac{M_{\rm pl}^2\Hub^2}{\epsilon}(\delta_M-12\epsilon+ 8 \delta)
C_0^{(\epsilon)},\nonumber\\
\ea
and for the mean field equation 
\ba
&&\bigg[1+\frac{8C_0^{(\epsilon)}}{\epsilon}
\bigg]\bphi''
+2\Hub\bigg[1-\frac{4C_0^{(\epsilon)}}{\epsilon}\bigg]\bphi'
+a^2V_{,\phi}\nonumber\\
&&\qquad\qquad\qquad\qquad\qquad=
-\frac{M_{\rm pl}^2C_0^{(\epsilon)}}{\epsilon}a^2V_{,\phi\phi\phi}[\bphi].
\ea
where the $(\epsilon)$ indicates keeping only leading order in slow-roll. Inserting the explicit expressions for the correlators, we find 
\ba
\label{eq:Fri1}
3M_{\rm pl}^2\Hub^2\bigg[1
-\frac{\Hub^2}{M_{\rm pl}^2}\frac{(\delta_M-12\epsilon) N_{\rm eff}}{24 \pi^2 a^2}
\bigg]&=&\left[\frac{1}{2}\bphi'^2+a^2V[\bphi]\right],\nonumber\\
\\
\label{eq:Fri2}
3M_{\rm pl}^2\Hub'\bigg[1-
\frac{\Hub^2}{M_{\rm pl}^2}\frac{(\delta_M-12\epsilon+8\delta)N_{\rm eff}}{24 \pi^2 a^2}
\bigg]
&=&\left[-\bphi'^2+a^2V[\bphi]\right]
,\nonumber\\
\ea
and 
\ba
\label{eq:FieldQ}
\bigg[1+\frac{\Hub^2}{M_{\rm pl}^2}\frac{N_{\rm eff}}{\pi^2  a^2}\bigg]\bphi''
+&2\Hub\bigg[1-\frac{
\Hub^2}{M_{\rm pl}^2}\frac{ 4 N_{\rm eff}}{8\pi^2 a^2}\bigg]\bphi'
+a^2V_{,\phi}\nonumber\\
&=
-\frac{\Hub^2 N_{\rm eff}}{8 \pi^2 a^2}a^2V_{,\phi\phi\phi}[\bphi]\,.\quad
\ea
 We saw that $\delta_M$ is really a slow-roll parameter, and so to leading order in SR (\ref{eq:Fri1}) reduces to the tree-level expression, whereas (\ref{eq:Fri2}) does not. Both agree with the tree-level expression in the semi-classical {\it limit} $H\ll M_{\rm pl}$. The field equation (\ref{eq:FieldQ}) has corrections on the LHS that vanish in the semi-classical {\it limit}; and a correction on the RHS that survives in both the semi-classical and SR limits. 

In the strict sense of slow-roll, $\bphi''$ is sub-leading in the field equation and $(\bphi')^2$ is sub-leading in the Friedmann equations, and should be discarded at this order. On the other hand, one may adopt the view that only quantum corrections should be truncated in slow-roll because the modes are, and they are small corrections to the full field/Friedmann equations. 

\section{The semi-classical approach}
\label{sec:semi-classical}

Our aim is to make the connection to the semi-classical approach, which we therefore present here. 
The prescription of semi-classical gravity is to treat gravitational degrees of freedom classically
 (no fluctuations) and the matter field(s) quantum mechanically (with fluctuations) in the induced curved
 spacetime.
This amounts to explicitly setting $A=\psi=B=E=0$ from the beginning, while retaining $\delta\phi$ as non-zero. As mentioned, this is in principle ambiguous, since by a gauge transformation, the four gravitational variables mix with $\delta\phi$. 

But following the semi-classical reasoning, taking the expectation value of the field operator and truncating at some order in connected correlators gives an approximation to the full field dynamics. We find, in the analogous 1-loop truncation as for the complete calculation
\ba
3M_{\rm pl}^2\Hub^2&=&
\frac{1}{2}\bphi'^2+\frac{1}{2}\langle\delta\phi'^2\rangle+\frac{1}{2}\langle(\nabla\delta\phi)^2\rangle\nonumber\\
&&
+a^2V[\bphi]+\frac{1}{2}a^2 
V_{,\phi\phi}[\bphi]\langle\delta\phi^2\rangle,
\\
3M_{\rm pl}^2\Hub'&=&-\bphi'^2-
\langle\delta\phi'^2\rangle+\nonumber\\
&&
a^2V[\bphi]+\frac{1}{2}a^2 V_{,\phi\phi}[\bphi]\langle\delta\phi^2\rangle,
\\
0&=&\bphi''+2\Hub\bphi'+a^2V_{,\phi}[\bphi]+\frac{1}{2}a^2V_{,\phi\phi\phi}[\bphi]\langle\delta\phi^2\rangle.\nonumber\\
\ea
There is a separate equation for the quantum mode functions,
\ba
\delta\phi_{\bf k}''+2\Hub\delta\phi_{\bf k}'+\left[{\bf|k|}^2+a^2V_{,\phi\phi}[\bphi]\right]\delta\phi_{\bf k}=0,
\ea
which is identical to (the Fourier transform of) Eq. (\ref{eq:fluct}) for $\psi=A=0$.

The quantisation of the fluctuations is no longer constrained, the commutation relations are simply 
\ba
\frac{1}{a^2}[\delta\phi(\mathbf{x}),\Pi_{\delta\phi}(\mathbf{y})]=[\delta\phi(\mathbf{x}),\delta\phi'(\mathbf{y})]
 = \frac{i}{a^2}\delta^3(\mathbf{x}-\mathbf{y})\,,\nonumber\\
\ea
and
\ba
 [{\hat{a}}_\mathbf{k},{\hat{a}}^\dagger_{\mathbf{k}'}]=(2\pi)^3 \delta^3(\mathbf{k}-\mathbf{k}')\,,
\ea
with all other commutators vanishing. This is indeed the $\epsilon\rightarrow 0$ or $\Hub / |{\bf k}|
 \rightarrow 0$ limit of (\ref{commutation_dphi2}). Using a similar definition to Eq.(\ref{field_operator})
\ba
\delta\phi(\eta,\mathbf{x}) =  \int \frac{d^3 k}{(2\pi)^3 a}
\left[a_\mathbf{k} h_\mathbf{k}(\eta) e^{i \mathbf{k \cdot x}} + a_\mathbf{k}^{\dagger} h_\mathbf{k}^*(\eta)
 e^{-i \mathbf{k \cdot x}}\right]\,,\nonumber\\
\ea
The Wronskian obeys the familiar
\ba
h_\mathbf{k}(\eta) h_\mathbf{k}'^*(\eta) -h_\mathbf{k}^*(\eta) h_\mathbf{k}'(\eta) =i,
\ea
and the mode equation reduces to
\ba
h_\mathbf{k}''(\eta)+\left[-\Hub^2-\Hub'+{\bf|k|}^2+a^2V_{,\phi\phi}[\bphi]\right]h_\mathbf{k}(\eta)=0.\nonumber\\
\ea
Inserting again $\Hub=-(1+\epsilon)/\eta$ and $\epsilon = 1-\Hub'/\Hub^2$, and taking again $\delta_M$ (\ref{eq:deltaM}) to be small and constant, we have the equation to leading order in slow-roll
\ba
h_\mathbf{k}''(\eta)+\left[\frac{-2-3\epsilon+\delta_M}{\eta^2}+|{\bf k}|\right]h_\mathbf{k}(\eta)=0.
\ea
The solution is
\ba
\label{SC_mode}
h_\mathbf{k}(\eta)=\sqrt{\frac{\pi}{4}}(-\eta)^{1/2}H^{(2)}_\nu(k\eta),
\nonumber\\
\nu=\frac{3}{2}\sqrt{1+\frac{4}{9}(3\epsilon-\delta_M)}\simeq \frac{3}{2}+\epsilon-\frac{\delta_M}{3}.
\ea
The $C_{0,2}$ correlators can be computed analogously to Section \ref{sec:loop} (see Appendix
 \ref{app:loop_integrals}) and we present the result in
 Appendix \ref{app:SC_correlators}. Ignoring UV divergences and keeping only leading order in slow-roll
 we have
\ba
\tilde{C}_0&\equiv&\langle\delta\phi^2\rangle=\frac{\Hub^2}{4\pi^2a^2}N_{\rm eff}^{\rm SC} ,
\nonumber\\
\tilde{C}_2&\equiv&\langle(\nabla\delta\phi)^2\rangle=-\frac{\Hub^4}{8\pi^2 a^2}\Lambda_{\rm IR}^2\simeq 0.
\label{eq:SC_correlators}
\ea
We have defined
\begin{eqnarray}
\label{NeffSC}
N_{\rm eff}^{\rm SC} \equiv \frac{1}{2(\delta_M / 3 - \epsilon)}
\bigg(1 - e^{-2(\delta_M / 3 - \epsilon)|\log\Lambda_{\rm IR}|}\bigg) \nonumber\\\simeq \left\{ \begin{array}{ll}
N\,, & \qquad N \ll N_{\rm sat}\,,\\
N_{\rm sat}\,, & \qquad N \gg N_{\rm sat}\,,\quad\delta_M > 3\epsilon\,,\\
N_{\rm sat}e^{N/N_{\rm sat}}\,, & \qquad N \gg N_{\rm sat}\,,\quad\delta_M < 3\epsilon\,,
\end{array}\right.
\end{eqnarray}
with
\begin{equation}
N_{\rm sat}^{\rm SC} \equiv \frac{1}{2|\delta_M / 3 - \epsilon|}\,.
\end{equation}
We then find
\ba
\frac{\partial_\eta\tilde{C}_0}{\Hub}=\frac{\partial_\eta^2\tilde{C}_0}{\Hub
^2}=\bigg(-2\epsilon+\frac{\partial_\eta N_{\rm eff}^{\rm SC}}{\Hub N_{\rm eff}^{\rm SC}}\bigg)\tilde{C}_0\equiv (-2\epsilon+\xi )\tilde{C}_0,
\label{eq:SC_correlators2}\nonumber\\
\ea
where we have taken $\xi$ to be first order in SR, and $\partial_\eta \xi$ to be of higher order. 
The equations of motion are given by 
\ba
3M_{\rm pl}^2\Hub^2&=&\frac{1}{2}\bphi'^2+a^2V[\bphi]\nonumber\\
&&+\frac{\Hub^2}{4}\bigg[
\frac{\partial_\eta^2\tilde{C}_0}{\Hub^2}+
2\frac{\partial_\eta \tilde{C}_0}{\Hub}+
4\delta_M\tilde{C}_0+
4\frac{\tilde{C}_2}{\Hub^2}\bigg]
,\nonumber\\
\\
3M_{\rm pl}^2\Hub'&=&-\bphi'^2+
a^2V[\bphi]\nonumber\\
&&-\frac{\Hub^2}{2}\bigg[
\frac{\partial_\eta^2\tilde{C}_0}{\Hub^2}
+2\frac{\partial_\eta \tilde{C}_0}{\Hub}
+\delta_M\tilde{C}_0
+2\frac{\tilde{C}_2}{\Hub^2}\bigg]
,\nonumber\\
\\
0&=&\bphi''+2\Hub\bphi'+a^2V_{,\phi}[\bphi]+\frac{1}{2}a^2V_{,\phi\phi\phi}[\bphi]\tilde{C}_0.\nonumber\\
\ea
Finally, inserting the explicitly leading terms in slow-roll Eqs.~(\ref{eq:SC_correlators}, \ref{eq:SC_correlators2}), we find 
\ba
\label{eq:Fri1SC}
3M_{\rm pl}^2\Hub^2\bigg[1
-\frac{\Hub^2}{M_{\rm pl}^2}\frac{(2\delta_M-3\epsilon+\frac{3}{2}\xi)N_{\rm eff}^{\rm SC}}{24\pi^2a^2}
\bigg]=\frac{1}{2}\bphi'^2+a^2V[\bphi]
,\nonumber\\
\\
\label{eq:Fri2SC}
3M_{\rm pl}^2\Hub'\bigg[1
+\frac{\Hub^2}{M_{\rm pl}^2}\frac{(\delta_M-6\epsilon+3\xi)N_{\rm eff}^{\rm SC}}{24\pi^2a^2}
\bigg]=-\bphi'^2+a^2V[\bphi]
,\nonumber\\
\\
\label{eq:FieldSC}
\bphi''+2\Hub\bphi'+a^2V_{,\phi}[\bphi]=-\frac{\Hub^2N_{\rm eff}^{\rm SC}}{8\pi^2a^2}a^2V_{,\phi\phi\phi}[\bphi].\nonumber\\
\ea
Comparing directly with (\ref{eq:Fri1}, \ref{eq:Fri2}, \ref{eq:FieldQ}), we see that the Friedmann equations again have corrections on the LHS,  similar to but not identical to the complete calculation. 

However,  these correction are again suppressed in the semi-classical limit $H\ll M_{\rm pl}$. Conversely, for the field equation the corrections on the LHS present in (\ref{eq:FieldQ}) are absent in the semi-classical approach. But the correction on the RHS is of the same form. 
Moreover, since by relation (\ref{eq:deltaM}) $\delta - 2\epsilon \simeq \delta_M / 3 - 3\epsilon$ we find that the saturation values $N_{\rm sat}$ and $N_{\rm sat}^{\rm SC}$ do not in general agree, however, in the perturbative regime $N \ll N_{\rm sat}$ we see that $N_{\rm eff} = N_{\rm eff}^{\rm SC} = N$ and hence the corresponding quantum corrections agree with the results of the full calculation.  

We conclude that the two calculations agree in the strict semi-classical limit: $H / M_{\rm pl} \to 0$ in the perturbative regime $N \ll N_{\rm sat}$, but that the leading corrections in $H^2 / M_{\rm pl}^2$ do not agree, and the leading corrections in SR also not. 

 In the strict semi-classical limit, the Friedmann equations reduce to the tree-level ones, and the field equation is simply (\ref{eq:FieldSC}). We note that this is simply the "Hartree" term, where we have inserted the (not Hartree resummed) slow-roll modes into the correlator, giving the leading $\Hub^2$ behaviour. Hence, during and shortly after inflation, the leading quantum correction is not the Minkowski space vacuum contribution, as sometimes advocated in the literature. 

The SC equations of motion (\ref{eq:Fri1SC}-\ref{eq:FieldSC}) are found to agree with the 1PI results of 
\cite{Herranen:2013raa} up to a finite renormalization, since in the
 saturated regime we can write $(2\delta_M-3\epsilon)N_{\rm sat}^{\rm SC} = \delta_M N_{\rm sat}^{\rm SC} + 3/2$
 and $(\delta_M-6\epsilon)N_{\rm sat}^{\rm SC} = - \delta_M N_{\rm sat}^{\rm SC} + 3$ in the Friedman equations
 (\ref{eq:Fri1SC}, \ref{eq:Fri2SC}), respectively, and further absorb the constants $3/2$ and $3$
 (times $H^4$) into the counter terms of the higher order gravity operators\footnote{In dimensional
 regularization where the counter terms of gravity operators can be chosen covariantly these involve $R^2$,
 $R^{\mu\nu}R_{\mu\nu}$ and $R^{\mu\nu\rho\sigma}R_{\mu\nu\rho\sigma}$.} as discussed in \cite{Herranen:2013raa}.

\section{Magnitude of radiative corrections: Examples}
\label{sec:examples}

In this section we estimate the magnitude of quantum radiative corrections in the mean field equation
 (\ref{eq:FieldQ}) for a couple of popular single-field inflation models.

\subsection{Large-field monomial inflation}
\label{sec:phi4}

Let us first consider the case
\ba
V[\phi]= \frac{\lambda}{24}\phi^4,
 \ea
 with $\phi$ slow-rolling from an initial condition $\phi\gg M_{\rm pl}$.
 In the slow roll limit we may neglect the second time-derivative in the mean field equation
 (\ref{eq:FieldQ}) to get
\ba
\label{eq:FieldQ_SR}
2\Hub\bigg[1-\frac{\Hub^2}{M_{\rm pl}^2}\frac{ N_{\rm eff}}{2\pi^2 a^2}\bigg]\bphi'
+a^2V_{,\phi}= -\frac{\Hub^2 N_{\rm eff}}{8 \pi^2 a^2}a^2V_{,\phi\phi\phi}[\bphi].\nonumber\\
\ea
By assuming that the quantum corrections are small we may solve (\ref{eq:FieldQ_SR}) iteratively to zeroth
 order: $\bphi' = -a^2 V_{,\phi} / (2 \Hub)$, and plug this back to the first order (in quantum corrections)
 term on the LHS of (\ref{eq:FieldQ_SR}) to get\footnote{
 Corrections to this iterative manipulation of the mean field equation (\ref{eq:FieldQ_SR}) would be of
 second order in slow roll and quantum corrections.}:
\ba
\label{eq:FieldQ_SR_iter}
2\Hub\bphi' + a^2V_{,\phi} = -\bigg[\frac{\Hub^2N_{\rm eff}}{6\cdot2\pi^2 a^2 }
\frac{\bphi^2}{M_{\rm pl}^2} + \frac{\Hub^2 N_{\rm eff}}{8 \pi^2 a^2}\bigg]a^2\lambda\bphi.\nonumber\\
\ea
We see that the non-SC quantum correction term (first term on the RHS) is related to the SC term 
 by the factor:
\ba
\frac{2}{3}\frac{\bphi^2}{M_{\rm pl}^2} \approx \frac{16}{3\epsilon} \approx \frac{16}{3}(N_e + 1) \sim 10^2\,-\,10 ^3,
\ea
where $N_e$ denotes the number of $e$-foldings from the end of inflation. So we find, rather surprisingly,
 that for $\phi^4$ inflation the non-SC quantum correction appears to be dominant in comparison to the SC
 quantum correction. Apparently during inflation the semi-classical limit $H/M_{\rm pl}\ll1$ is not sufficiently realised.

Moreover, based on Eqs.~(\ref{Neff}, \ref{spectrum_rel}), we find that for a realistic value of the scalar spectral index, $n_s < 1$, including the metric perturbations the quantum radiative corrections would grow exponentially large for
 $N \gg N_{\rm sat}$ signalling the breakdown of perturbative expansion for $N \gtrsim N_{\rm sat}$ \footnote{In the SC, there is no such breakdown as $\delta_M/3-\epsilon>0$; but the non-SC correction is then absent. }. 
 Assuming that the true non-perturbative saturation value for $N_{\rm eff}$ would be of order $N_{\rm sat} \approx N_e$,
 we can estimate the relative size of the quantum corrections by comparing the (dominant)
 correction term inside the square brackets on the LHS of (\ref{eq:FieldQ_SR}) to unity to find an upper
 limit 
\ba
\frac{\Hub^2}{M_{\rm pl}^2}\frac{N_{\rm eff}}{2\pi^2 a^2} \lesssim  \frac{4\lambda N_e^3}{3\pi^2}
\lesssim 10^5 \lambda \sim 10^{-7}\,,
\label{eq:phi4_est}
\ea
i.e. even the saturated value of the IR enhanced quantum correction would be negligibly small compared to
 the tree-level terms in the mean field equation. In the last relation in (\ref{eq:phi4_est}), we have used a typical value for the self-coupling consistent with the CMB.
 
\subsection{Small-field inflation}
\label{sec:inverted}

Consider instead a hill-top small-field inflation model:
\ba
V[\phi]=\Lambda^4\left(1 - \frac{\phi^4}{\mu^4} + \ldots \right),
\ea
with the understanding that $\phi=0$ initially, and subsequently slow-rolls to the bottom of the potential
 determined by the higher order terms. Following
 the above steps we find that the non-SC correction in the mean field equation (\ref{eq:FieldQ}) is related to
 the SC correction by the same factor:
\ba
\frac{2}{3}\frac{\bphi^2}{M_{\rm pl}^2} \ll 1\,,\qquad {\rm for} \qquad \mu \ll M_p\,, 
\ea
which, however, is small in this case and hence the SC correction dominates over the non-SC correction,
 as expected for a small-field inflation model. As before, for a realistic value $n_s < 1$ 
the perturbative expansion breaks for $N \gtrsim N_{\rm sat}$. If we again assume for the true
 non-perturbative saturation value $N_{\rm eff} \sim N_{\rm sat} \sim N_e$, we find for the relative size of the dominant
 SC quantum correction in comparison to tree-level terms:
\ba
\frac{\Hub^2}{\bar\phi^2}\frac{6 N_{\rm eff}}{8\pi^2 a^2} \lesssim \frac{9 A_s}{4 N_e} \sim 10^{-10}\,,
\ea
where $A_s$ denotes the amplitude of the primordial scalar perturbations in CMB, with $A_{s*} \approx 2.22\times 10^{-9}$
 for $k_* = 0.05\;{\rm Mpc}^{-1}$. We find that in this case the quantum radiative corrections are even smaller than
 for the previous large-field example, and so although the semi-classical limit is enforced and a SC computation is valid, the resulting corrections are negligible. 

\subsection{Exponential inflation}
\label{sec:exponential}

Finally, consider the following exponential potential:
\ba
V[\phi] = \frac{\lambda M_{\rm pl}^4}{4\xi^2}\left(1 - \exp\left(-\frac{2\phi}{\sqrt{6}M_{\rm pl}}\right)\right)^2\,,
\ea
corresponding to Higgs inflation in the Einstein frame \cite{Bezrukov:2007ep}. Following the above steps we now find
 that the non-SC quantum correction in the mean field equation (\ref{eq:FieldQ}) is larger by a factor of 6 compared to
 the SC correction term. Again, for a realistic value $n_s < 1$  
the perturbative expansion breaks for $N \gtrsim N_{\rm sat}$, and assuming that for the true
 non-perturbative saturation value $N_{\rm eff} \sim N_{\rm sat} \sim N_e$, we now find for the relative size of the dominant
 non-SC quantum correction:
\ba
\frac{\Hub^2}{M_{\rm pl}^2}\frac{N_{\rm eff}}{2\pi^2 a^2} \lesssim  \frac{\lambda N_e}{24\pi^2 \xi^2}
\sim 10^{-9}\,,
\label{eq:Higgs_est}
\ea
where we have used $\xi \simeq 47000\sqrt{\lambda}$ following from the CMB normalization. Again we find that even the
 fully saturated radiative corrections are tiny in comparison to tree level contributions.

\section{Discussion and Conclusion}
\label{sec:conclusion}

In the standard calculation, the evolution of the background cosmological degrees of freedom $H$ and $\bphi$ determine the spectrum of density perturbations through the slow-roll parameters at horizon exit. To leading order in slow roll the scalar spectral index and tensor-to-scalar ratio are given by
\ba
\label{spectrum}
n_s = 1 - 6\epsilon + \frac{2}{3}\delta_M\,,\qquad\qquad r = 16\epsilon\,,
\ea
respectively. At the level of our approximation (quadratic action in fluctuations) the quantum radiative corrections are accounted for in the mean field and Friedmann equations but not in the mode functions, or more generally, in the 2-point functions. Therefore, in this approximation, the expressions for the power spectra do not involve explicit quantum corrections and Eq.~(\ref{spectrum}) remains formally intact. The quantum radiative corrections to these observables then enter through the corrections in the mean field and Friedman equations, affecting the values of $\epsilon$ and $\delta_M$ at horizon crossing.    

In the present work, we have computed the leading loop order, leading order in slow-roll radiative corrections to the mean field and Friedmann evolution equations, both in the semi-classical approach and including the metric scalar
fluctuations. 
We found that the two give qualitatively different results, but that they both reduce to the same in the semi-classical {\it limit}, in the perturbative regime. One is however not the limit of the other, and away from the semi-classical limit and/or the perturbative regime (when the saturated values of $N_{\rm eff}$ apply), the two calculations diverge. 

In the evaluation of the loop integrals the IR cutoff was related to the number of $e$-folds from the beginning of
 inflation by $|\log\Lambda_{\rm IR}| \approx N$, corresponding to a prescription where the modes that have been 
 outside the horizon for the whole duration of inflation are cut off. For the observed value of the scalar
 spectral index, $n_s < 1$, we then found that the IR contributions grow exponentially large for $N \gtrsim N_{\rm sat}
 = 1/|n_s - 1|$, signalling the breakdown of perturbative expansion and calling for a non-perturbative resummation.
 Such a Hartree resummation in the SC approach was carried out in \cite{Herranen:2013raa}, and similar studies using
 the non-perturbative stochastic approximation \cite{Riotto:2008mv,Starobinsky1,Starobinsky2,Garbrecht1,Garbrecht2} indicate that for large
 $N$ the loop contributions indeed saturate after linear growth (as a function of $N$) to a value depending on the model. In order to carry out such a resummation program for the full coupled system including the metric
 perturbations, one would have to expand the Lagrangian at least to cubic order in fluctuations.
 
To be able to estimate the size of quantum radiative corrections for realistic models in the present work, we assumed
 that the true non-perturbative saturation value for the loop contributions would be determined approximately by extrapolating
 the linear growth until the breaking point ($N \sim N_{\rm sat}$). Interestingly, we found that for large-field
 inflation models considered in Sections \ref{sec:phi4} and \ref{sec:exponential} there is no window where the semi-classical approach is a good approximation, since the non-SC $H/M_{\rm pl}$-suppressed terms dominate over the SC corrections. In addition, these radiative corrections are very small for realistic values of the couplings due to the CMB normalisation constraint.

On the other hand, for the small-field model of Section \ref{sec:inverted} the SC corrections dominate over the non-SC ones by virtue of $\bar\phi \ll M_{\rm pl}$ and therefore in this case the SC approach provides a solid approximation between the full calculation including the metric scalar perturbations and the classical tree-level calculation. However, also in this case the quantum radiative corrections are tiny for realistic values of parameters.

In more general theories, where the couplings are not all forced to be small to allow for inflation (multiple field models, curvaton models, ...), the overall size of radiative corrections may not be negligible, and our work suggests that using the SC {\it approach} does not in general provide the semiclassical {\it limit} of the complete calculation. Resolving IR issues through resummations in the former does therefore not imply a solid approximation to the latter, unless deep in the semi-classical {\it limit}. { By choosing the gauge appropriately, only the fluctuations of one field need mix with the metric perturbations, and additional fields may therefore explicitly be treated semi-classically, in the sense described here.} This is the subject of ongoing work. At this point, the prudent approach to IR artefacts seems to be resummation in the complete calculation; this means including diagrams to all orders in the couplings, where also tensor perturbation may play a role. 

\vspace{0.5cm}

\noindent {\bf Acknowledgments}: 
We would like to thank Tommi Markkanen for enlightening discussions and collaboration on related topics. AO thanks Gerasimos Rigopoulos for useful discussions. MH and AT are supported through a Villum Foundation Young Investigator Grant.

\appendix

\section{Quadratic action in the Newtonian gauge}
\label{app:action}
In the Newtonian gauge $E = B = 0$ the quadratic action (up to second order in the fluctuations $\{\delta\phi$, $A$, $\psi\}$) is given by 
\ba
\label{action_app}
S= S_{\rm g}+S_{\rm \phi}\,,
\ea
where the part involving the scalar field can be written as (in conformal time $\eta$, $dt=a(\eta)d\eta$,
 derivatives denoted by $'$)
\ba
\label{matter_action_app}
S_{\phi}=S_\phi^{(0)}+S_\phi^{(1)}+S_\phi^{(2)},
\ea
with
\ba
\label{eq:expaction1}
&&S_{\phi}^{(0)}=\int d^3x\,d\eta\, a^2\bigg[\frac{1}{2}\eta^{\mu\nu}\partial_\mu\bphi\,\partial_\nu\bphi-a^2V[\bphi]\bigg],\\
&&S_\phi^{(1)}
\left[\phi,g_{\mu\nu}\right]=\nonumber\\
&&\int d^3x\,d\eta \,a^2\bigg[(A-3\psi)\left(\frac{1}{2}\eta^{\mu\nu}(\partial_\mu\bphi)(\partial_\nu\bphi)-a^2V[\bphi]\right)
\nonumber\\
&&\qquad-A(\partial_0\bphi)(\partial_0 \bphi)
+\left(\eta^{\mu\nu}(\partial_\mu \delta\phi)(\partial_\nu \bphi)-\delta\phi\, a^2V_{,\phi}[\bphi]\right)\bigg],\nonumber\\
\\
&&S_\phi^{(2)}\left[\phi,g_{\mu\nu}\right]=\nonumber\\
&&\int d^3x\,d\eta\, a^2\bigg[\left( 
-\frac{A^2}{2}-3A\psi +\frac{3\psi^2}{2}\right)\times
\nonumber\\&&\qquad\qquad\qquad\qquad\qquad\qquad
\left(\frac{1}{2}\eta^{\mu\nu}(\partial_\mu\bphi)(\partial_\nu \bphi)-a^2V[\bphi]\right)
\nonumber\\&&
\qquad+\left(\frac{1}{2}\eta^{\mu\nu}(\partial_\mu \delta\phi)(\partial_\nu \delta\phi)-\frac{1}{2}\delta\phi^2 \,a^2V_{,\phi\phi}[\bphi]\right)
\nonumber\\&&
\qquad+2A^2(\partial_0\bphi)(\partial_0\bphi)-A\,\partial_0\delta\phi\,
\partial_0\bphi
-A(A-3\psi)(\partial_0\bphi)(\partial_0\bphi)
\nonumber\\
&&
\qquad+(A-3\psi)(\eta^{\mu\nu}(\partial_\mu\delta\phi)(\partial_\nu\bphi)-\delta\phi \,a^2V_{,\phi}[\bphi])
\bigg]\,,\nonumber\\&&
\label{eq:expaction2}
\ea
while the strictly gravitational part is given by
\ba
S_{\rm g}=S_{\rm g}^{(0)}+S_{\rm g}^{(1)}+S_{\rm g}^{(2)},
\ea
with
\ba
S_{gr}^{(0)}&=&\frac{1}{16\pi G}\int d^3x\,d\eta \,a^{2}6(\mathcal{H}'+
\mathcal{H}^2),
\\
S_{gr}^{(1)}&=&-\frac{1}{16\pi G}\int d^3x\,d\eta
\,a^2\bigg[ -6\mathcal{H}^2A+6\,\psi
(2\mathcal{H}'+\mathcal{H}^2)
\bigg],\nonumber\\
\\
S_{gr}^{(2)}&=&\frac{1}{16\pi G}\int
d^3x\,d\eta \,a^2\bigg[-6\psi'^2-12
\mathcal{H}(A+\psi)\psi'
\nonumber\\
&&\qquad\qquad-9\mathcal{H}^2
(A+\psi)^2
-2\psi_{,i}(2A_{,i}-\psi_{,i})\bigg].\nonumber\\
\label{last_eq}
\ea

\section{Equations of motion}
\label{app:eom}
Variation of the action w.r.t a generel metric leads to the Einstein equation
\ba
E_{\nu}^\mu=R_{\nu}^\mu-\frac{1}{2}\delta^\mu_\nu R=8\pi GT_\nu^\mu.
\ea
We insert the perturbed metric Eq. (\ref{eq:pert_met}) in Newtonian gauge and compute the Christoffel symbols, Ricci tensor, Ricci scalar. These are then combined into
\ba
a^2 E^0_0&=&3\mathcal{H}^2\left(1-2A+4A^2\right)
+2(1+4\psi)\nabla^2\psi+3(\nabla\psi)^2\nonumber\\
&&-3\nabla\psi\nabla A
+3
\left(\psi'-2\mathcal{H}
-4\psi\mathcal{H}
+4\mathcal{H}A\right)\psi',
\\
a^2E^i_j&=&
\bigg(2\mathcal{H}'+\mathcal{H}^2-8\mathcal{H}[\psi\psi'-A\psi'-AA']-\psi'[\psi'+2A']
\nonumber\\
&&-2\psi''[1+2(\psi-A)]
+4[2\Hub'+\mathcal{H}^2]
A^2
-2\mathcal{H}(A'+2\psi')\nonumber\\
&&
-2(2\mathcal{H}'+
\mathcal{H}^2)A
+\nabla^2(\psi-A)-2(\psi-A)\nabla^2A\nonumber\\
&&+3\nabla\psi\nabla A+(\nabla A)^2+4\psi\nabla^2\psi+2(\nabla\psi)^2
\bigg)\delta^i_j
\nonumber\\
&&-[(1+2\psi)(\psi-A)_{,ij}-A_{,i}\psi_{,j}-A_{,j}\psi_{,i}+
2\psi\psi_{,ij}\nonumber\\
&&+3\psi_{,i}\psi_{,j}
+
A_{,j}A_{,i}+2AA_{,ij}].\nonumber\\
\ea
The energy momentum tensor is calculated by inserting the perturbed metric and  
\ba
\phi&=&\bar{\phi}+\delta\phi,\\
 V[\phi]&=&V[\bar{\phi}]+\delta
\phi V[\bar{\phi}]_{,\phi}+
\frac{1}{2}\delta\phi^2V[\bar{\phi}]
_{,\phi\phi}
\ea
into the expression
\ba
T^\mu_\nu&=\phi^{,\mu}\phi_{,\nu}
-\left[\frac{1}{2}\phi^{,\alpha}\phi_{,\alpha}-V[\phi]\right]
\delta^\mu_\nu.
\ea
This gives
\ba
a^2T^0_0
&=&\frac{1}{2}\bar{\phi}'^2+a^2V[
\bar{\phi}]
-\bar{\phi}'^2A(1-2A)
+\bar{\phi}'\delta\phi'(1-2A)\nonumber\\
&&+
\delta\phi \,a^2V_{,\phi},
+\frac{1}{2}
\delta\phi'^2
+\frac{1}{2}(\nabla\delta\phi)^2+\frac{1}{2}\delta\phi^2a^2V_{,\phi\phi},
\nonumber\\
a^2T^i_j
&=&\bigg(-\frac{1}{2}\bar{\phi}'^2+a^2V[\bar{\phi}]
+\bar{\phi}'^2A(1-2A)
-\bar{\phi}'\delta\phi'(1-2A)\nonumber\\
&&
+
\delta\phi \,a^2V_{,\phi}
-\frac{1}{2}\delta\phi'^2+\frac{1}{2}(\nabla\delta\phi)^2
+\frac{1}{2}\delta\phi^2\,a^2V_{,\phi\phi}
\bigg)\delta^i_j\nonumber\\
&&
-\delta\phi_
{,i}\delta\phi_{,j},
\ea
 with the linear equation (\ref{eq:fluct}) for the fluctuation $\delta\phi$.
To linear order in perturbations, the $i\neq j$ component of the energy momentum tensor vanishes. Equating this to the expression for $ E^i_j$  $i\neq j$, also to linear order, we conclude that $ A=\psi $ . 
We note that when we later take vacuum expectation values of the equations, terms linear in the perturbations will vanish. 
Inserting the perturbed metric and splitting the field into a background value $ \bar{\phi} $ and a fluctuation, $ \delta\phi $ we get the following result to second order
\ba
&&\partial_\mu\left(a^{-2}\sqrt{-g}a^2g^{\mu\nu}\partial_
\nu\phi\right)+\sqrt{-g} V_{,\phi}=0
\nonumber\\&&
\qquad \Rightarrow
{\bar{\phi}}''+
2\mathcal{H}
{\bar{\phi}}'+
a^2V,_{\bar{\phi}}(\bar{\phi})=
-6\bar{\phi}''\psi^2\nonumber\\
&&\qquad\qquad
-12\bar{\phi}'[2
\mathcal{H}
\psi
+\psi']\psi
+\delta\phi''4\psi
+4[2\mathcal{H}\psi+\psi']\delta\phi'
\nonumber\\
&&\qquad\qquad+
2\psi^2a^2V_{,\bar{\phi}}
+2\psi\delta\phi \,a^2V_{,\bar{\phi}\bar{\phi}}
-\frac{1}{2}\delta\phi^2 a^2V_{,\bar{\phi}
\bar{\phi}\bar{\phi}}.\nonumber\\
\ea
 In total we obtain the set of equations (\ref{eq:fried1}-\ref{eq:fried4}) for the gravitational fluctuations at linear order.
Finally, computing the $ ^0_i $ Einstein equation to first order gives  the following relation
\ba
\psi'+\mathcal{H}A&=4\pi G\bar{\phi}'\delta\phi.
\ea
Inserting slow roll parameters and writing terms quadratic in the perturbation in terms of $ C_0,C_2,C_4 $ as described in appendix \ref{sec:correlators} enables us to combine the Einstein equations to (4.3) and (4.4) and write the field equation as (4.1).

\section{Correlator relations}
\label{sec:correlators}
The equations of motion in Appendix \ref{app:eom} include terms linear and quadratic in the fluctuations $\psi$, and through constraint relations, $A$ and $\delta \phi$. Taking the quantum expectation of these equations in the vacuum state,  the linear terms in fluctuations vanish, by the definition of the separation into $\bphi+\delta\phi$, and $\bar{g}_{\mu\nu}$ and $A$, $\psi$, $E$, $B$. At quadratic order, we define the expectation values to be the symmetric ones, so that for instance
\ba
\langle\psi'\psi\rangle&\rightarrow& \frac{1}{2}\langle \psi'\psi + \psi\psi'\rangle,\nonumber\\
\langle\psi''\psi\rangle&\rightarrow& \frac{1}{2}\langle \psi''\psi + \psi\psi''\rangle,\nonumber\\ &\qquad& \ldots,
\ea
with the understanding that all correlators are equal-time and -space, with this limit to be taken after any differentiation is performed.
By taking further time-derivatives and using the mode equation (\ref{mode_eq}) we find the
 following relations, which will be useful to us in the following:
\begin{align}
\frac{1}{2}\langle \psi'\psi + \psi\psi'\rangle &= \frac{1}{2}\partial_\eta \langle \psi^2 \rangle,
\nonumber\\
\frac{1}{2}\langle \psi''\psi + \psi\psi''\rangle &= -\mathcal{H}\delta\partial_\eta \langle \psi^2 \rangle
-2\mathcal{H}^2(\delta-\epsilon) \langle \psi^2 \rangle + \big\langle \psi\nabla^2 \psi \big\rangle,
\nonumber\\
\langle \psi'^2 \rangle &= \frac{1}{2}\partial_\eta^2 \langle \psi^2 \rangle
 - \frac{1}{2}\langle \psi''\psi + \psi\psi''\rangle,
\nonumber\\
\frac{1}{2}\langle \psi''\psi' + \psi'\psi''\rangle &= -2\mathcal{H}\delta\langle \psi'^2 \rangle
-\mathcal{H}^2(\delta-\epsilon) \partial_\eta \langle \psi^2 \rangle
 + \frac{1}{2}\partial_\eta \big\langle \psi\nabla^2 \psi \big\rangle,
\nonumber\\
\frac{1}{2}\langle \psi'''\psi + \psi\psi'''\rangle &= -\left[(\mathcal{H}\delta)' 
- 2\mathcal{H}^2\delta^2 + \mathcal{H}^2(\delta-\epsilon)\right]\partial_\eta \langle \psi^2 \rangle
\nonumber\\
&
 + \frac{1}{2}\partial_\eta \big\langle \psi\nabla^2 \psi \big\rangle - 2\mathcal{H}\delta \big\langle \psi\nabla^2 \psi \big\rangle
\nonumber\\
&- 2 \left[\big(\mathcal{H}^2(\delta-\epsilon)\big)' - 2\mathcal{H}^3\delta(\delta-\epsilon)\right]
\langle \psi^2 \rangle
,
\nonumber\\[3mm]
\langle \psi''^2\rangle &= 4\mathcal{H}^2\delta^2 \langle \psi'^2 \rangle
+ 4\mathcal{H}^3\delta(\delta-\epsilon) \partial_\eta \langle \psi^2 \rangle
\nonumber\\
&
+ 4\mathcal{H}^4(\delta-\epsilon)^2 \langle \psi^2 \rangle ,
- 2\mathcal{H}\delta\partial_\eta \big\langle \psi\nabla^2 \psi \big\rangle
\nonumber\\
&- 4\mathcal{H}^2(\delta-\epsilon) \big\langle \psi\nabla^2 \psi \big\rangle
+ \big\langle \psi\nabla^4 \psi \big\rangle\,.
\label{correlator_relations}
\end{align}
By these relations we can express all the loop contributions in the background equations of motion in terms
 of $C_0=\langle \psi^2 \rangle$, $C_2=\big\langle \psi\nabla^2 \psi \big\rangle$ and
 $C_4=\big\langle \psi\nabla^4 \psi \big\rangle$ given by Eqs.~(\ref{loop_final}-\ref{k4loop_final}) and their
 $\eta$-derivatives.

\section{Commutation relations}
\label{app:comm_relations}
In this section we present the equal-time commutation relations for $\psi$, $\delta\phi$ and their
 conjugate momenta $\Pi_\psi$, $\Pi_{\delta\phi}$. Using (\ref{constr_matrix}-\ref{equal-time_comm_def})
we get by a straightforward calculation:
\footnote{Identical commutation relations would have been recovered by including all four
constraints $\chi_1,\ldots,\chi_4$ in the ($4 \times 4$) constraint matrix $C_{ij}$.}
\begin{align}
&[\psi(\mathbf{x}),\psi(\mathbf{y})] = [\Pi_\psi(\mathbf{x}),\Pi_\psi(\mathbf{y})] = 0,
\nonumber\\[1mm]
&[\psi(\mathbf{x}),\Pi_\psi(\mathbf{y})] = i\frac{12 \pi G \bphi'^2}{\nabla_x^2}
\delta^3(\mathbf{x}-\mathbf{y})\,,
\label{commutation_psi}
\end{align}
\begin{align}
&[\delta\phi(\mathbf{x}),\delta\phi(\mathbf{y})]
 = [\Pi_{\delta\phi}(\mathbf{x}),\Pi_{\delta\phi}(\mathbf{y})] = 0,
\nonumber\\[1mm]
&[\delta\phi(\mathbf{x}),\Pi_{\delta\phi}(\mathbf{y})]
 = i\left(1 - \frac{12 \pi G \bphi'^2}{\nabla_x^2}\right)
\delta^3(\mathbf{x}-\mathbf{y})\,,
\label{commutation_dphi}
\end{align}
and
\begin{align}
&[\psi(\mathbf{x}),\delta\phi(\mathbf{y})]
 = - i\frac{4 \pi G \bphi'}{a^2 \nabla_x^2}\delta^3(\mathbf{x}-\mathbf{y}),
\nonumber\\
&[\Pi_\psi(\mathbf{x}),\Pi_{\delta\phi}(\mathbf{y})]
 = - i 3 a^2\big(\bphi' + \frac{3\mathcal{H}' \bphi'
 + a^2 \mathcal{H} V_{,\phi}}{\nabla_x^2}\big)\delta^3(\mathbf{x}-\mathbf{y}),
\nonumber\\
&[\psi(\mathbf{x}),\Pi_{\delta\phi}(\mathbf{y})]
 = i\frac{4\pi G \left(3 \mathcal{H} \bphi' + a^2 V_{,\phi}\right)}{\nabla_x^2}
\delta^3(\mathbf{x}-\mathbf{y}),
\nonumber\\
&[\delta\phi(\mathbf{x}),\Pi_\psi(\mathbf{y})]
 = - i\frac{3 \mathcal{H} \bphi'}{\nabla_x^2}\delta^3(\mathbf{x}-\mathbf{y})\,,
\label{commutation_off-diag}
\end{align}
where we have suppressed the (equal) time arguments of the fields. The various commutation
 relations involving $\delta\phi$ can also be obtained by solving
 $\delta\phi$ and $\Pi_{\delta\phi}$ from the constraints
 $\chi_3$ and $\chi_2$ and using Eqs.~(\ref{commutation_psi}) for the commutation relations
 of $\psi$ and $\Pi_\psi$ and therefore the commutation algebra (\ref{commutation_dphi}
-\ref{commutation_off-diag}) is consistent with all the constraints.

\section{Calculation of one-loop correlators}
\label{app:loop_integrals}

To evaluate the one-loop integral (\ref{loop_text})
\begin{align}
\label{loop}
\langle \psi^2 \rangle  =  \int \frac{d^3 k}{(2\pi)^3} |f_\mathbf{k}|^2
 = \frac{\epsilon \Hub^2}{16\pi M_p^2 a^2}
\int_{\Lambda_{\rm IR}}^{\Lambda_{\rm UV}} dx \big|H^{(2)}_{\nu}(-x)\big|^2\,,
\end{align}
we notice that the Hankel functions $H^{(1,2)}_{\nu}(x)$ have a branch cut along
 the negative real axis $x < 0$ and here we pick the branch by the relation
 $H^{(2)}_{\nu}(e^{-i\pi}x) = e^{i\pi\nu}H^{(1)}_{\nu}(x)$ such that for $x > 0$
\begin{align}
\big|H^{(2)}_{\nu}(-x)\big|^2 = \big|H^{(1)}_{\nu}(x)\big|^2\,.
\end{align}
We then separate the momentum integration in Eq.~(\ref{loop}) in three parts:
\begin{align}
\int_{\Lambda_{\rm IR}}^{\Lambda_{\rm UV}} = \int_{\Lambda_{\rm IR}}^{\kappa_{\rm IR}} + \int_{\kappa_{\rm IR}}^{\kappa_{\rm UV}}
+ \int_{\kappa_{\rm UV}}^{\Lambda_{\rm UV}}\,,
\end{align}
with
\begin{align}
 \Lambda_{\rm IR} \ll \kappa_{\rm IR} \ll 1 \ll \kappa_{\rm UV} \ll \Lambda_{\rm UV}\,.
\end{align}
The low-momentum (IR) part of the loop integral can be evaluated by using the $x \to 0$
 asymptotic expansion of the Hankel function:
\begin{align}
H_{\nu}^{(1)}(x)=-i\bigg(\frac{2}{x}\bigg)^\nu\frac{\Gamma[\nu]}{\pi}+\mathcal{O}(x^\nu)\,,
\end{align}
to get
\begin{align}
\label{IR_cont}
&\int_{\Lambda_{\rm IR}}^{\kappa_{\rm IR}} dx\,\big|H^{(1)}_{\nu}(x)\big|^2
= \frac{2^{2\nu} \Gamma^2(\nu)}{\pi^2(1-2\nu)}\left(\kappa_{\rm IR}^{1-2\nu} - \Lambda_{\rm IR}^{1-2\nu} \right)
\nonumber\\
&= \frac{2}{\pi}\left[\bigg(\frac{1}{2(\delta-2\epsilon)} - \frac{1}{2} + \log 2 + \gamma_E\bigg)
\bigg(1 - \Lambda_{\rm IR}^{2\delta - 4\epsilon}\bigg) + \log\kappa_{\rm IR}\right]
\nonumber\\
 &+ {\cal O}(\epsilon,\delta)\,.
\end{align}
For the intermediate-momentum contribution we may set $\epsilon,\delta \to 0$ and use
\begin{align}
H_{1/2}^{(1)}(x)=-i\sqrt{\frac{2}{\pi x}}e^{i x},
\end{align}
to get
\begin{align}
\label{int_cont}
\int_{\kappa_{\rm IR}}^{\kappa_{\rm UV}} dx\,\big|H^{(1)}_{\nu}(x)\big|^2
= \frac{2}{\pi}\Big[\log\kappa_{\rm UV} - \log\kappa_{\rm IR}\Big] + \mathcal{O}(\epsilon,\delta)\,.
\end{align}
Finally, Using the large-$|x|$ asymptotic expansion:
\begin{align}
H^{(1)}_\nu(x)&=-\frac{e^{i \left(x-\frac{\pi  \nu }{2}\right)}}{\sqrt{\pi  x}}\bigg[{(1 - i) }
+\frac{(1 +i) \left( \nu ^2-1/4\right)}{2 x}
\nonumber\\
&-\frac{(1 - i)\left(9-40 \nu ^2+16 \nu ^4\right)}{128x^2}
\bigg]
+\mathcal{O}(x^{-7/2})\,,
\end{align} 
we get for the high-momentum (UV) part
\begin{align}
\int_{\kappa_{\rm UV}}^{\Lambda_{\rm UV}} dx\,\big|H^{(1)}_{\nu}(x)\big|^2
 = \frac{2}{\pi}\Big[\log\Lambda_{\rm UV} - \log\kappa_{\rm UV}\Big] + \mathcal{O}(\epsilon,\delta)\,.
\end{align}
By combining these expressions we get for the loop integral (\ref{loop}) in total
\begin{align}
C_0\equiv\langle \psi^2 \rangle &= \frac{\epsilon \Hub^2}{8\pi^2 M_p^2 a^2}
\bigg[\bigg(\frac{1}{2(\delta-2\epsilon)} - \frac{1}{2} + \log 2 + \gamma_E\bigg)\times\nonumber\\
&\bigg(1 - \Lambda_{\rm IR}^{2\delta - 4\epsilon}\bigg) + \log\Lambda_{\rm UV}\bigg]
+ {\cal O}(\epsilon,\delta)\,.
\end{align}
Similarly, we get for the $\mathbf{k}^2$-loop:
\begin{align}
C_2\equiv&-\big\langle \psi\nabla^2 \psi \big\rangle = \int \frac{d^3 k}{(2\pi)^3} \mathbf{k}^2 |f_\mathbf{k}|^2
\nonumber\\
&= \frac{\epsilon \Hub^2}{16\pi^2 M_p^2 a^2}(-\eta)^{-2}
\Big[\Lambda_{\rm UV}^2 -(\delta - 2\epsilon)\log\Lambda_{\rm UV} - \Lambda_{\rm IR}^2\Big]
+ \ldots\,,
\end{align}
and for the $\mathbf{k}^4$-loop:
\begin{align}
C_4\equiv&\big\langle \psi\nabla^4 \psi \big\rangle = \int \frac{d^3 k}{(2\pi)^3} \mathbf{k}^4 |f_\mathbf{k}|^2
\nonumber\\
&= \frac{\epsilon \Hub^2}{32\pi^2 M_p^2 a^2}(-\eta)^{-4}
\Big[\Lambda_{\rm UV}^4 - (\delta - 2\epsilon)\Lambda_{\rm UV}^2\nonumber\\& + \frac{1}{16}(\delta - 2\epsilon)^2
(2 + \delta - 2\epsilon)^2\log\Lambda_{\rm UV}
- \Lambda_{\rm IR}^4\Big]
\nonumber\\ &\qquad\qquad\qquad\qquad\quad+ \ldots\,.
\end{align}
\section{Correlators in the semi-classical approach}
\label{app:SC_correlators}
In the semi-classical approach the mode functions are given by (\ref{SC_mode}). 
Computing the analog of the $C_{0,2}$ correlators, we find (see section \ref{sec:loop} for the notation)
\ba
\tilde{C}_0&=&\int \frac{d^3\mathbf{k}}{(2\pi)^3}\frac{1}{a^2}|h_\mathbf{k}|^2=\frac{1}{8\pi^2 a^2}(-\eta)^{-2}\int^{\Lambda_{\rm UV}}_{\Lambda_{\rm IR}} dx\, x^2|H^{(1)}_\nu(x)|^2=\nonumber\\
&&
\frac{1}{4\pi^2a^2}(-\eta)^{-2}\bigg[\left(\frac{1}{2(\delta_M/3-\epsilon)}+\frac{\log 2}{2}+\psi(3/2)\right)\times
\nonumber\\
&&\left(1-\Lambda_{\rm IR}^{2(\delta_M/3-\epsilon)}\right)+\frac{\Lambda_{\rm UV}^2}{2}+\log\Lambda_{\rm UV}\bigg]
+\mathcal{O}(\epsilon, \delta_M).\\
\tilde{C}_2&=&\int \frac{d^3\mathbf{k}}{(2\pi)^3}\frac{{\bf k}^2}{a^2}|h_\mathbf{k}|^2=\frac{1}{8\pi^2 a^2}(-\eta)^{-4}\int^{\Lambda_{\rm UV}}_{\Lambda_{\rm IR}} dx\, x^4|H^{(1)}_\nu(x)|^2\nonumber\\&=&\frac{1}{8\pi^2 a^2}(-\eta)^{-4}\left[\frac{1}{2}\Lambda_{\rm UV}^4+\Lambda_{\rm UV}^2-\Lambda_{\rm IR}^2\right],
\ea
and we also have the relation
\ba
\langle (\delta\phi')^2 \rangle &= \frac{1}{2}\partial_\eta^2 \langle \delta\phi^2 \rangle
 - \frac{1}{2}\langle \delta\phi''\delta\phi + \delta\phi\delta\phi''\rangle\nonumber\\&
 =\frac{1}{2}\partial_\eta^2\tilde{C}_0+
 \mathcal{H}\partial_\eta\tilde{C}_0
 +\mathcal{H}^2\delta_M\tilde{C}_0+
 \tilde{C}_2.     
\ea

\end{document}